\documentstyle[11pt,newpasp,twoside,epsf]{article}
\def\edcomment#1{\iffalse\marginpar{\raggedright\sl#1\/}\else\relax\fi}
\reversemarginpar

\begin{document}
\title{XMM-Newton search for hot gas in the dwarf galaxy IC~2574}
\author{Michael Kappes, J\"urgen Kerp}
\affil{Radioastronomisches Institut der Universit\"at Bonn, Auf dem H\"ugel 71, 53121 Bonn, Germany}
\author{Fabian Walter}
\affil{National Radio Astronomy Observatory, P.O. Box O, Socorro, NM, 87801, USA}

\begin{abstract}
We present an XMM--Newton observation of the nearby dwarf galaxy IC~2574. This galaxy is part of our ensemble observed within a multi--frequency campaign to study the physical conditions of the interstellar medium in dwarf galaxies in general.

Here, we focus on the diffuse soft ($0.2 - 0.5\,{\rm keV}$) X--ray emission associated with IC~2574. Since this emission is expected to be quite faint we have to reach the true X--ray background intensity level. For this aim, we have to account for unrelated diffuse X--ray emission and for the vignetting of the X--ray mirrors aboard the XMM--Newton satellite.
\end{abstract}

%----------------------------------------------Introduction
\section{Introduction}
\label{intro}
According to the standard model the interstellar medium of dwarf galaxies is thought to be strongly shaped by stellar evolution. Supernovae and/or stellar winds produce holes and bubbles in the interstellar medium of their host galaxies. The linear extent of these structures reach $\sim 1\,\rm{kpc}$ sizes. To illustrate this, Fig. 1 shows the \ion{H}{i} distribution of IC~2574 measured by the VLA. It is very likely that hot gas --- produced by stellar activity --- is expelled into the halo of the galaxy because the gravitational potential of low--mass dwarf galaxies is very shallow. Therefore, we expect to find hot coronal gas in the vicinity of the galaxy body. In the case of IC~2574 these hypothetical plasma structures are expected on a tens--of--arcmin scale. Henceforward, the XMM--Newton satellite with its 30\arcmin\ FOV and angular resolution of 14\arcsec\ (half energy width) is the ideal instrument to detect coronal gas in dwarf galaxies.

\begin{figure}
\plotfiddle{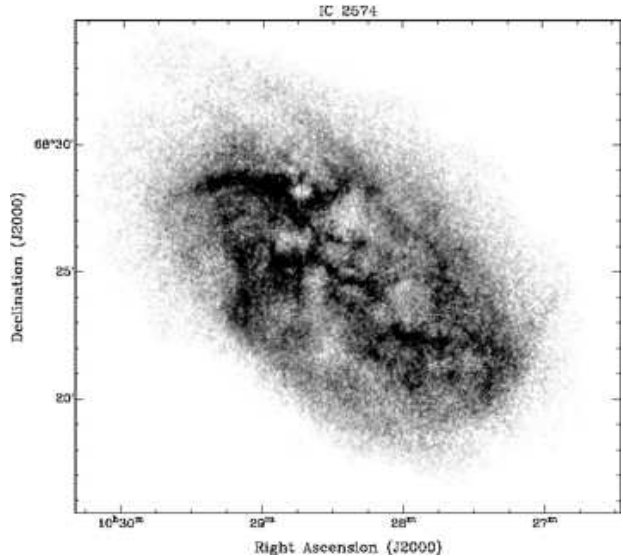}{7.1cm}{0}{90}{90}{-130}{0}
\caption{\ion{H}{i} distribution of IC~2574 measured by the Very Large Array (VLA). A lot of sub--structure shows up as holes and bubbles on an arcmin scale. They are thought to be produced by stellar activity (stellar winds, supernovae).}
\end{figure}

%----------------------------------------------Systematics
\section{Systematic effects of XMM--Newton data}
\label{sys}
The diffuse soft X--ray emission is a very faint signal in comparison to the contribution of X--ray point sources. Therefore, the reduction of the X--ray data is a crucial point.

%----------------------------------------------Flarefilter
\subsection{Solar proton--flares}
\label{flare}
The EPIC instruments aboard XMM--Newton are affected by proton--flares originating from the Sun. These proton--flares show up as peaks in the lightcurves (total counts vs. exposure time) of the EPIC data, producing an artificial count rate across the whole FOV, which is by a factor of up to $\sim10$ higher than the 'true' X--ray signal. It is essential to flag the time intervals affected by proton--flares. For this aim, we developed an algorithm which rejects the flares on a user definable $\sigma$--level.

We compute the mean ($\mu_1$) of the lightcurve and flag the photon counts which exceed a certain threshold (depending on the chosen $\sigma$--level). Then the new mean ($\mu_2$) is calculated --- which is smaller than $\mu_1$ --- and the photons exceeding the adapted $\sigma$--level are flagged as well. The Iteration stops when $\mu_1-\mu_2 < \sqrt{\mu_2}$\,, i.e. the statistical uncertainty of the data.

The flagged time intervals are excluded from further analysis. This is the first major step in preparing the XMM--Newton EPIC data for the investigation of diffuse soft X--ray emission.

%----------------------------------------------Vignetting
\subsection{Vignetting}
\label{vig}
Like all imaging X--ray telescopes the XMM--Newton EPIC instruments show the effect of {\em vignetting}. This is the reduction of effective area which depends on the photon energy and off--axis angle. The effect is clearly visible as a gradual decrease of sensitivity towards the edges of the FOV. 
It is possible to correct for this effect by calculating a so--called {\em exposure map}. Using the SAS\,5.3.3 {\tt expmap} task results in exposure maps which are inconsistent with the observational data. This is exemplified in Fig. 2. The SAS exposure maps lead to an over--correction of the data even in the center region of the FOV.
\begin{figure}
\plotone{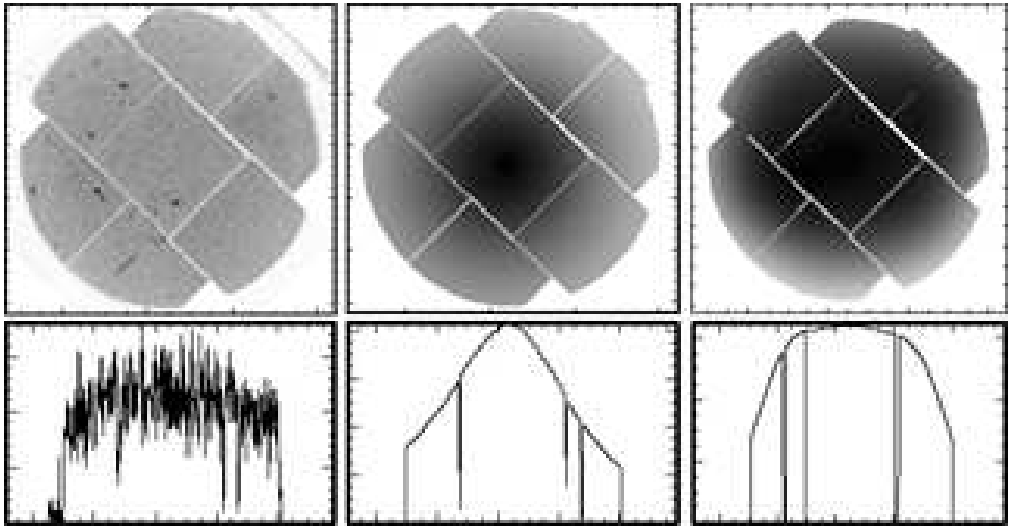}
\caption{{\bf left:} The XMM--Newton deep field north is one of the deep pointings we used to calculate an improved exposure map. The profile below the image is a horizontal slice through the FOV center. Vignetting produces the gradual fading towards the edges.
{\bf middle:} The exposure map as calculated with SAS\,5.3.3. The profile below shows also the vignetting which is too steep to match the observational data even in the central region of the FOV (left).
{\bf right:} Our improved exposure map as calculated from deep observations ($>400\,{\rm ks} total$). The profile below shows also the expected vignetting effect but is more flat in the center of the FOV. This matches much better the observational situation.
}
\end{figure}
In order to correct for the vignetting we produce improved exposure maps for the EPIC MOS instruments using five deep observations ($>400\,\rm{ks}$ total) obtained from the XMM--Newton Science Archive. Combining the individual deep field data is sufficient to sample the full FOV and moreover, it minimizes the contribution of point sources. This approach is more confident than the {\tt expmap} task because we make use of the XMM--Newton inflight performance. After the calibration of the deep field data we remove all point sources down to a flux limit of $4.5 \times 10^{-15} \rm{erg\,s^{-1} cm^{-2}}$ (in the energy range $0.2 - 0.5\,{\rm keV}$) using the {\tt eboxdetect} and {\tt emldetect} tasks. These yield so--called {\em cheese images} which show holes at the point source positions. Since the SAS calculated cut--out radii are too small, we developed software which calculates the correct radii and fills up the holes with a statistical photon noise map including the intensity normalization for each hole. After combining (i.e. adding) all 'hole filled' images of the multiple deep pointings for each energy band, we smooth these 'merged' exposure maps and multiply by the detector mask. The resulting exposure map is displayed in Fig. 2 (right). 

%----------------------------------------------IC2574
\section{Results and Outlook}
\label{ic}
The flare--filter algorithm and the improved exposure maps are applied to the calibrated XMM--Newton data of IC~2574. After filtering the lightcurves and dividing the images by the exposure maps, both MOS\,1 and MOS\,2 are combined to achieve a better S/N. The brightest point sources down to $\approx 6 \times 10^{-4}\,{\rm cts\,s^{-1}}$ are removed and the image is adaptively smoothed envoking {\tt csmooth} of the CIAO\,2.2.1 software package. The resulting image in the energy range of $0.2 - 0.5\,{\rm keV}$ is shown in Fig. 3.
\begin{figure}
\plotfiddle{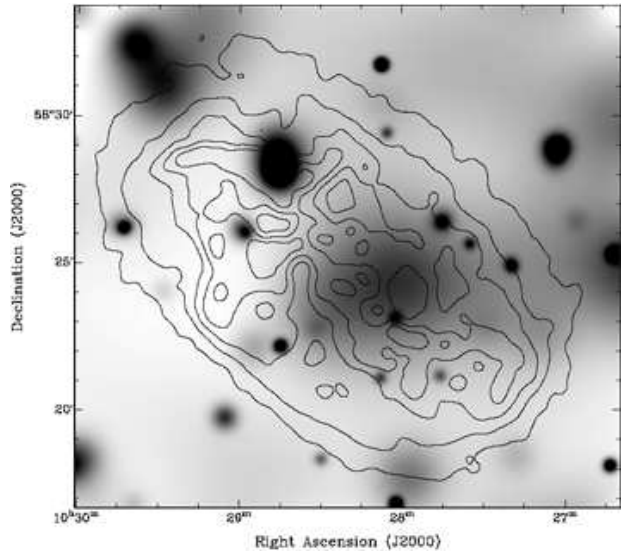}{7.1cm}{0}{90}{90}{-130}{0}
\caption{Adaptively smoothed image of the combined point source subtracted MOS\,1/MOS\,2 data ($0.2 - 0.5\,{\rm keV}$) and {\ion{H}{i}} contours. After removing the brightest point sources the diffuse soft X--ray emission is detected near the center of the map.}
\end{figure}

%We can reject the solar proton--flares very accurate providing a maximum exposure time. The exposure map correction works pretty well and is more confident than the procedure used in the SAS\,5.3.3 package, because we use observational data for computing the exposure map.

With this technique the detection limits of the IC~2574 observation for point sources and diffuse emission are ${\rm L_x}\approx 1.5 \times 10^{36} {\rm erg\,s^{-1}}$ and ${\rm L_x}\approx 4 \times 10^{37} {\rm erg\,s^{-1}}$ respectively. It turns out that diffuse soft X--ray emission emerges from the galaxy. Since the emission is visible in a very soft energy window ($0.2 - 0.5\,{\rm keV}$) it is likely that the emission originates from in front of the \ion{H}{i} distribution. Otherwise it would not be detectable due to the photoelectric absorption of the \ion{H}{i}. However, a better S/N can be achieved when the EPIC PN instrument is included.
\end{document}